\documentclass[12pt,preprint,apjfonts]{emulateapj}
\usepackage{psfig}
\slugcomment{Accepted by ApJLett}

\newcommand{\nc}{\newcommand}

\nc{\be}[1]{\begin{equation}\mbox{$\label{#1}$}}
\nc{\bea}[1]{\begin{eqnarray} \mbox{$\label{#1}$}}
\nc{\Section}[2]{\section{#2}\label{#1}}
\nc{\Bibitem}[1]{\bibitem{#1}}
\nc{\Label}[1]{\label{#1}}

\nc{\Mpc}{Mpc/h}
\nc{\vev}[1]{\langle #1 \rangle}

\nc{\eea}{\end{eqnarray}}
\nc{\ee}{\end{equation}}
\nc{\eeq}{\end{equation}}

\def\lcdm{$\Lambda$CDM~}

\def\ltsima{$\; \buildrel < \over \sim \;$}
\def\gtsima{$\; \buildrel > \over \sim \;$}
\def\simlt{\lower.5ex\hbox{\ltsima}}
\def\simgt{\lower.5ex\hbox{\gtsima}}
\nc{\w}{$w_2(\theta)$\ }
\nc{\ie}{i.e.} 
\nc{\eg}{e.g.}
\def\q{{\hat n} }

\shorttitle{Detection of the ISW and SZ effects}
\shortauthors{Fosalba et al.}

\begin{document}

\title{Detection of the ISW and SZ effects 
from the CMB-Galaxy correlation}

\author{Pablo Fosalba$^{1}$, Enrique Gazta\~{n}aga$^{2,3}$, Francisco J
Castander$^{2}$ } 

\altaffiltext{1}{Institut d'Astrophysique de Paris, 98bis Bd Arago,
75014 Paris, France} 
\altaffiltext{2}{Institut d'Estudis Espacials de Catalunya/CSIC, Gran
Capit\`a 2-4, 08034 Barcelona, Spain}
\altaffiltext{3}{INAOE, Astrofisica, Tonantzintla, Puebla 7200, Mexico}

\begin{abstract}

We present a cross-correlation analysis of the WMAP cosmic microwave
background (CMB) temperature anisotropies and the SDSS galaxy density
fluctuations.  We find significant detections of the angular
CMB-galaxy correlation for both a flux limited galaxy sample (z $\sim$
0.3) and a high redshift (z $\sim$ 0.5) color selected sample. The
signal is compatible with that expected from the integrated
Sachs-Wolfe (ISW) effect at large angles ($\theta>4^\circ$) and the
Sunyaev-Zeldovich (SZ) effect at small scales ($\theta<1^\circ$).  The
detected correlation at low-z is in good agreement with a previous
analysis using the APM survey ($z\sim 0.15$).  The combined analysis
of all 3 samples yields a total significance better than 3$\sigma$ for
the ISW and about $2.7$$\sigma$ for the SZ, with a Compton parameter
$\overline{y}\simeq 10^{-6}$.  For a given flat \lcdm model, the ISW
effect depends both on the value of $\Omega_\Lambda$ and the galaxy
bias $b$. To break this degeneracy, we estimate the bias using the
ratio between the galaxy and mass auto-correlation functions in each
sample. With our bias estimation, all samples consistently favor a
best fit dark-energy dominated model: $\Omega_\Lambda \simeq 0.8$,
with a 2$\sigma$ error $\Omega_\Lambda=0.69-0.86$.

\end{abstract}

\keywords{cosmic microwave background, cosmology: observations}

\section{Introduction}
\label{sec:intro}

A recent study \citep{fosalba:2003} has cross-correlated the cosmic
microwave background (CMB) anisotropies measured by WMAP
\citep{bennett/etal:2003} with galaxy fluctuations in the APM Galaxy
Survey \citep{maddox/etal:1990} to find significant detections for
both the integrated Sachs-Wolfe (ISW) and thermal Sunyaev-Zeldovich
(SZ) effects.  The ISW detection is in agreement with other analyses
based on X-ray and radio sources \citep{boughn/crittenden:2003,
nolta/etal:2003},
while \citet{hernandez/rubino:2003} fail to detect the SZ effect when
comparing WMAP to different cluster templates (see also
\citet{myers:2003}).
It should be stressed nevertheless that cluster
or galaxy group catalogues are too sparse and typically produce worse
signal-to-noise ratios than galaxy surveys.  Moreover, depending on
the sample  there could be a significant
cancellation of the ISW and SZ effects on scales smaller than a few
degrees (see \S\ref{sec:pred}).
In this letter we cross-correlate the WMAP CMB temperature
anisotropies with galaxies from the  Sloan Digital
Sky Survey (SDSS; \citealt{york/etal:2000}).
When we were finishing this work we became aware of a similar
analysis \citep{scranton/etal:2003}  
that uses different color and photometric redshift selected 
samples from the SDSS.

\section{Data}
\label{sec:data}

We make use of the largest datasets currently available to study the
CMB-galaxy cross-correlation. In order to probe the galaxy
distribution, we have selected subsamples from the SDSS Data
Release 1 (SDSS DR1; \citealt{abazajian/etal:2003}) which covers $\sim
2000$ deg$^2$ (i.e, 5 $\%$ of the sky).  The samples analyzed here
have different redshift distributions and a large number of galaxies
(10$^5$-10$^6$, depending on the sample). We concentrate our analysis
on the North sky ($\sim$ 1500 deg$^2$, ie, 3.6 $\%$ of the sky),
because it contains the largest and wider strips. The South SDSS DR1
($\sim$ 500 deg$^2$) consists of 3 narrow and disjoint $2.5^\circ$
strips, which are less adequate for our analysis.

Our main sample, hereafter {\it SDSS all}, includes all objects
classified as galaxies with extinction corrected magnitude $r < 21$,
and a low associated error ($< 20 \%$). This sample contains $\sim$ 5
million galaxies distributed over the North sky. Its
predicted redshift distribution is broad and has a median redshift 
$\overline{z}\sim 0.3$. Our high-redshift sample ({\it SDSS high-z} thereafter)
comprises $\sim 3 \times 10^5$ galaxies, with $\overline{z}\sim 0.5$. 
It was selected by imposing magnitude cuts and color cuts perpendicular
to the redshift evolution and the spectral type
variations based on theoretical spectral synthesis models.
We shall also compare
our results to the APM analysis in \citet{fosalba:2003}, who used a
$b_J=17-20$ sample, $\overline{z}\simeq 0.15$, area $\sim 4300$ deg$^2$ and
1.2 million galaxies.

For the CMB data, we use the first-year full-sky WMAP maps
\citep{bennett/etal:2003}.  Since the observed CMB-galaxy correlation
is practically independent of the WMAP frequency band used
\citep{fosalba:2003}, we shall focus on the V-band ($\sim 61$ GHz) as
it conveniently combines low pixel noise and high spatial resolution,
$21^{\prime}$. In addition, we have also used the W-band and a
foreground ``cleaned'' WMAP map \citep{tegmark/etal:2003} to check that our results
are free from galactic contamination.  We mask out pixels
using the conservative Kp0 mask, that cuts out $21.4 \%$ of the sky
\citep{bennett/etal:2003}. All the maps used have been
digitized into $7^{\prime}$ pixels using HEALPix
\footnote{http://www.eso.org/science/healpix}, \citep{gorski/hivon/wandelt:1999}.


\section{Cross-Correlation and Statistical tests}
\label{sec:cross}

We follow the notation introduced in \citet{fosalba:2003}.
We define the cross-correlation function as the expectation value of density
fluctuations $\delta_G= N_G/<N_G>-1$ and temperature anisotropies
$\Delta_T= T-T_0$ (in $\mu$K)
at two positions $\q_1$ and $\q_2$ in the sky:
$w_{TG}(\theta) \equiv  \vev{ \Delta_T({\bf\q_1}) \delta_G({\bf\q_2}) }$,
where $\theta = |\bf{\q_2}-\bf{\q_1}|$. 

\begin{figure}
\centering{\vskip -2.5truecm 
\epsfysize=8cm \epsfbox{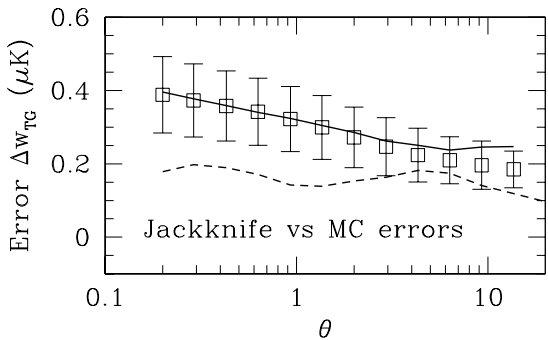}}
\caption{\label{fig:jackvs}
Errors in the cross-correlation $w_{TG}(\theta)$
from the dispersion in 200
Monte-Carlo simulations (solid line) as compared 
with the mean and dispersion (squares with errobars)
in the Jack-knife (JK) error estimation over the same simulations.
Dashed line correspond to the JK 
error in the real WMAP-{\it SDSS all} sample.}
\end{figure}

We compute the CMB-galaxy correlation and the associated statistical
error-bars using the jack-knife (JK) method described in
\citep{fosalba:2003} and references therein. The survey is divided
into $M=16$ (we find similar results for $M=8$) separate regions on
the sky, each of equal area. The $w_{TG}$ analysis is then performed
$M$ times, each time removing a different region, the so-called
JK subsamples. The covariance $C_{ij}$ for $w_{TG}$ between
scales $\theta_i$ and $\theta_j$ is obtained by re-scaling the
covariance of the JK subsamples by a factor $M-1$ (see Eq.[3]
in \cite{fosalba:2003}). To test the JK errors and covariance
we have also run $200$ WMAP V-band Monte-Carlo (MC)
realizations. We add random realizations of the measured WMAP
temperature angular power-spectrum \citep{bennett/etal:2003} 
to those of the white noise estimated for the relevant frequency band
\citep{hinshaw/etal:2003}. For each MC simulation we estimate the 
 mean ``accidental''correlation $w_{TG}$ of  
simulated CMB maps to the SDSS galaxy density fluctuation map.
We also estimate the associated JK error in each MC simulation.
Fig.\ref{fig:jackvs} compares the 'true' sampling error from the dispersion
of   $w_{TG}(\theta)$ in 200 MC simulations with the mean 
and dispersion of the JK errors
over the same simulations. The JK error gives an
excellent estimate of the 'true' error up  $\theta \simeq 5$ degrees.
On larger scales it only underestimates the error by $10-20\%$,
which is hardly significant given the  uncertainties.

\begin{figure}
\centering{\epsfysize=6.5cm \epsfbox{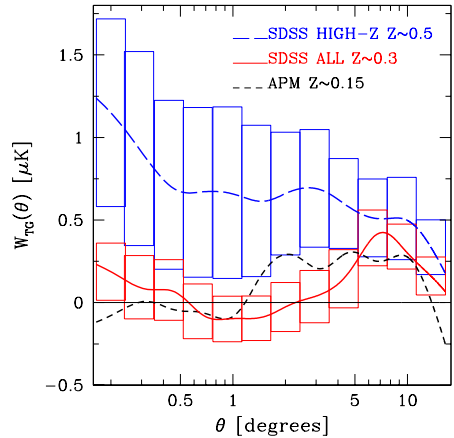}}
\caption{\label{fig:wtg_mc}
WMAP-SDSS correlation:
long dashed-line shows the measurement for the {\it SDSS HIGH-Z}
sample, while the solid line displays the correlation for the 
{\it SDSS ALL} sample. For reference, short-dashed line displays the same
measurement using the APM galaxy survey instead of SDSS.
Boxes show 1-$\sigma$ error-bars.}
\end{figure}

Fig.\ref{fig:wtg_mc} shows $w_{TG}$ for the different
samples together with the corresponding JK error.
It turns out that the  the JK errors from the real
WMAP sample  are in some cases smaller (up to a factor of two)
than the JK errors (or sample to sample dispersion) 
from the MC simulations.
Fig.\ref{fig:jackvs} shows, as a dashed-line,
 the comparison for the {\it SDSS all} sample,
which exhibits the largest discrepancy.
This difference in error estimation 
is not totally surprising as the MC simulations do
not include any physical correlations
and use a CMB power spectrum that is valid for the
whole sky, and not constraint as to match the CMB power over
the SDSS region. 
The JK errors provide a model free estimation that
is only subject to moderate ($20\%$) uncertainty, while MC errors
depend crucially on the model assumptions that go into the
simulations.
Despite these differences in the MC error estimation 
the overall significance for the detection turns out to
be smiliar, as explained in \S4.1.

We derive the significance of the detected correlation taking into
account the large (JK)
covariance between neighboring (logarithmic) angular
bins in survey sub-samples (but see also \S 4.1).  Adjacent bins at large scales
($\theta>4^{\circ}$) are correlated at the $\simeq 80\%$ level,
dropping to $\simeq 40\%$ for alternative bins.  Bins at smaller
scales are progressively more correlated.  To assign a conservative
significance for the detection (ie against $w_{TG}=0$) we estimate the
minimum $\chi^2$ fit for a constant $w_{TG}$ and give the difference
$\Delta \chi^2 $ to the $w_{TG}=0$ null detection. For example, at
scales  $\theta=4-10^{\circ}$ we find: $w_{TG}
= 0.53 \pm 0.21 \mu$K for the {\it SDSS high-z} sample, $w_{TG} = 0.26
\pm 0.13 \mu$K for the {\it SDSS all} sample and $w_{TG} = 0.35 \pm
0.13 \mu$K for the APM, in all cases we give 1-$\sigma$ errorbars.

We find the largest significance in the CMB-galaxy correlation for 
the {\it SDSS high-z} sample: $\Delta \chi^2=9.1$ (ie
probability, $P=0.3\%$ of no detection) for $\theta<10^\circ$ 
(being $\chi^2_{min}=14.6$ for  $w_{TG}=0.55 \mu$K with 11 d.o.f., 
although the fit is only approximate as the signal drops with scale).
In order to assess the
significance levels for the ISW and SZ effects 
from the observed CMB-Galaxy correlations,
we shall first introduce model predictions.

\section{Comparison to  Predictions}
\label{sec:pred}

\begin{figure}
\centering{\epsfysize=6.5cm \epsfbox{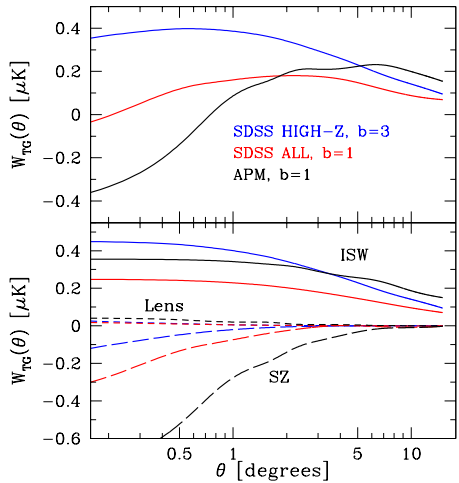}}
\caption{\label{fig:w2pre}
Theoretical predictions: 
(Bottom panel) Continuous, long and short dashed lines show the ISW, SZ
and lensing predictions. Different sets of lines correspond to the
APM (black), {\it SDSS all} (red) and {\it SDSS high-z} (blue)
samples. (Top panel) The total prediction (ISW+SZ+Lensing)
for the 3 samples. We have assumed a \lcdm model 
with a fixed $b_{gas}=2$ in all cases,
$b=3$ for {\it SDSS high-z} and $b=1$ for APM and {\it SDSS all}.}
\end{figure}

The temperature of CMB photons is gravitationally redshifted as they
travel through the time-evolving dark-matter gravitational potential
wells along the line-of-sight, from the last scattering surface $z_s =
1089$ to us, $z=0$ \citep{sachs/wolfe:1967}.  At a given sky
position ${\bf\q}$: $\Delta T^{ISW}({\bf\q}) = -2 \int
~dz~\dot{\Phi}({\bf\q},z)$, and 
for a flat universe $\nabla^2\Phi = -4\pi
G a^2 \rho_m \delta$ (see Eq.[7.14] in \citet{peebles:1980}). In
Fourier space it reads, $\Phi(k,z) = -3/2 \Omega_m
(H_0/k)^2\delta(k,z)/a$, and thus: 
\be{eq:wTG_ISW} w_{TG}^{ISW}(\theta)
= <\Delta_T^{ISW}\delta_G> = \int {dk\over k}~P(k)~g(k\theta) 
\ee 
being, $g(k\theta)={1/{2\pi}} \int dz~W_{ISW}(z)~W_G(z)~j_0(k\theta\,r)$,
where the ISW window is $W_{ISW} = -3\Omega_m({H_0/c})^2\dot{F}(z)$, 
with $c/H_0 \simeq 3000 h$ Mpc$^{-1}$,
$\dot{F}=d(D/a)/dr=(H/c)D(f-1)$, and $f \simeq
\Omega_m^{6/11}(z)$ quantifies the time evolution of the gravitational
potential.  The galaxy window function is $W_G \simeq
b(z)~D(z)~\phi_G(z)$, which depends on the galaxy bias, linear
dark-matter growth and the galaxy selection function. The ISW
predictions for the 3 samples are shown in in bottom panel of
Fig.\ref{fig:w2pre}. Unless stated otherwise, we use the concordance
$\Lambda$CDM model with $\Omega_m=0.3$, $\Omega_\Lambda=0.7$, $\Gamma
\simeq h\Omega_m =0.2$ and $\sigma_8=1$.

The weak lensing effect prediction is quite similar to the ISW,
we just need to replace the time derivative of the Newtonian potential by its
2D Laplacian \citep{SeljakZaldarriaga:2000}:
$W_{Lens} = 3k^2\Omega_m({H_0/c})^2(D/a)/d(r)$
$d(r)$ being the angular distance to the lensing 
sources (with: $d(r_s-r)/d(r_s) \simeq 1$).

For the thermal Sunyaev-Zeldovich (SZ) effect, we assume that the gas
pressure $\delta_{gas}$ fluctuations are traced by the galaxy
fluctuations $\delta_{gas} \simeq b_{gas}~\delta_G$ with a relative
amplitude given by the gas bias, $b_{gas} \simeq 2$, representative of
galaxy clusters, although $b_{gas}$ is uncertain to within 50 $\%$ on
linear scales and for low-z sources \citep{refregier/teyssier:2002}.
A simple conservative estimate of the SZ effect is given by
\citep{refregier/etal:2000}: 
\be{eq:sz} 
w_{TG}^{SZ}(\theta) = -b_{gas}~\overline{{\Delta T}} ~w_{GG}(\theta)
\ee
where
$\overline{{\Delta T}}$ is the mean temperature change in CMB photons
Compton scattered by electrons in hot intracluster gas.  Following
\citet{refregier/etal:2000}, we calculate $\overline{{\Delta T}} =
j(x) \overline{y} ~T_0$,  where $T_0 \simeq 2.73$K is the mean CMB temperature, 
$\overline{y}$ is the mean Compton
parameter induced by galaxy clusters, and $j(x) = -4.94$ is the
negative SZ spectral factor for the V-band.  The Compton parameter
can be calculated integrating along the line of sight the normalized
galaxy redshift distribution convolved with the volume-averaged
density-weighted temperature.  The latter is obtained from the mass
function and the M-T relation. We assume the Seth \&
Tormen mass function \citep{sheth/tormen:1999,sheth/etal:2001}
and the M-T relation given by
\citet{borgani/etal:1999}. In summary, we obtain for the WMAP V-band,
$\overline{{\Delta T}} = 6.65 \mu$K for the {\it SDSS all} sample, and
$\overline{{\Delta T}} = 6.71 \mu$K for the {\it SDSS high-z} sample
which corresponds to $\overline{y} \simeq
1.35 \times 10^{-6}$ for both samples.  The SZ predictions for the 3
samples are shown in the bottom panel of Fig.\ref{fig:w2pre}. Note
the galaxy auto-correlation explains most of the differences observed.

The total predicted correlation is thus the sum of
three terms: the ISW, thermal SZ and Lensing contributions, 
$w_{TG}=w_{TG}^{ISW}+w_{TG}^{SZ}+w_{TG}^{Lens}$.
Fig.\ref{fig:w2pre} shows individual contributions of these effects
(bottom panel) and the total (top) for the 3 samples analyzed. 
The ISW effect typically dominates for angles
$\theta >4^{\circ}$, while the SZ effect is expected to be significant
on small scales $\theta<1^{\circ}$. Lensing is found to be negligible
at all scales for our samples.

Before we can make a direct comparison between theory and observations,
we shall address the issue of galaxy bias.
The higher redshift sample
requires a high bias $b > 1$ to explain the large cross-correlation
seen at all scales
(the SZ effect being smaller at high redshift). At low redshifts 
the measured correlation is dominated by the thermal
SZ on small scales ($\theta<1^\circ$) and by ISW 
on large-scales ($\theta>4^\circ$). Here no bias is required to
match the observations. 
This agrees quite well with our self-consistent bias
estimation: for each sample we can estimate the ratio 
$b^2 \simeq w_{GG}/w_{MM}$, where
$w_{MM}$ and $w_{GG}$ are the (theoretically predicted)
matter and (measured) galaxy auto-correlation 
functions. For APM and {\it SDSS all} samples we find $b^2 \simeq 1$,
while for the {\it SDSS high-z} sample we get $b^2 \simeq 6$.

\subsection{Significance tests}
\label{sec:chi2}

\label{sec:isw}

\underline{\bf ISW effect:}
On large scales $\theta>4^\circ$, the ISW effect is expected to
dominate for all survey depths (see Fig \ref{fig:w2pre}).  Therefore,
from the large-angle CMB-Galaxy correlation, we can directly infer the
ISW effect (ie, $w_{TG}=w^{ISW}_{TG}$, see end of \S\ref{sec:cross}).  
In particular, for the {\it SDSS high-z} sample, a
constant correlation fit rejects the null detection with high
significance $\Delta \chi^2=6.0$ ($P=1.4\%$), comparable
to the level found for the APM survey, $\Delta \chi^2=6.1 ~(P=1.3\%)$.
A smaller significance is
obtained for the {\it SDSS all} sample: $\Delta \chi^2 =3.9$
(P$=4.8\%$). Alternatively, we can use the uncorrelated MC simulations (see \S3)
 to have an  independent estimate of the significance. 
When a particular MC simulation has an accidentally large
value of $w_{TG}$ it also has a large JK error associated.
We can thus assign a significance to our measurement by asking 
how many of the 200 MC simulations have a value of $w_{TG}$ 
equal or larger than the observations with an assosiated JK  error
equal or smaller than found for the observations. We find that only two
of the MC simulations fulfil this condition in any of the samples,
meaning that the significance of each detection 
is better than $1\%$ (for each of the 3 different data samples).

Since these samples are basically independent, we can combine them to
infer a total significance for the ISW detection: we find a total
$\Delta \chi^2=16$ ($P=0.1 \%$ for 3 d.o.f) corresponding to a 3.3$\sigma$.
Note we could do better using a (scale dependent) \lcdm model
theory prediction, but at the cost of introducing model dependent
detection levels.  Moreover, we can further include the ISW-dominated
small angle bins in our deepest sample, where SZ is negligible,
increasing the significance to $\Delta \chi^2=18.8$, 
($P=0.03\%$ for 3 d.of.), ie
we detect the ISW effect at the a 3.6$\sigma$ level.

\label{sec:sz}

\underline{\bf SZ effect:}
We can estimate the significance of the drop in the signal at small
angles in the {\it SDSS all} and APM samples due to the SZ effect (see Fig\ref{fig:w2pre})
using the best-fit constant at large angles (ie, the ISW signal) and ask for the observed
deviation from such value at smaller scales. For $\theta<1^{\circ}$, we find:
$w^{SZ}_{TG}=-0.27\pm 0.11$ for {\it SDSS all}, and 
$w^{SZ}_{TG}=-0.41\pm 0.16$ for APM (1-$\sigma$ errorbars).
Note this is conservative
because the ISW increases slightly as we approach smaller scales (see
\S IV). This test gives 
$\Delta \chi^2 =5.5 ~(P=2\%)$ for the {\it SDSS all} sample
and  $\Delta \chi^2 =8.5 ~(P=0.3\%)$ for the APM.

\section{Discussion}
\label{sec:discuss}

\begin{figure}
\centering{
\epsfysize=5.5cm \epsfbox{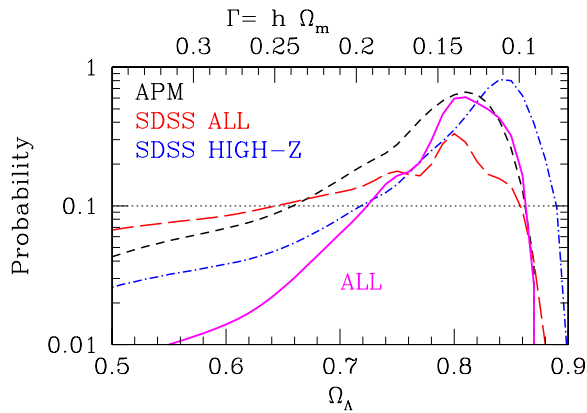}}
\caption{\label{fig:FitLambda}
Estimating dark-energy: Long-dashed, short-dashed and dot-dashed lines show the
probability distribution for $\Omega_\Lambda$ in the {\it SDSS all}, APM and
{\it SDSS high-z} samples. The combined distribution (for 3 d.o.f.) is shown by
the solid line.}
\end{figure}

We have measured the CMB-galaxy correlation using WMAP and
the SDSS DR1 galaxy survey. We measure a significant
cross-correlation at low (z $\sim 0.3$) and high (z $\sim 0.5$)
redshifts. We detect a positive correlation on large-scales induced by
the ISW effect at the 2$\sigma$ level for the (broadly distributed)
low-z sample.  This correlation is similar to that
measured for the lower redshift ($z \sim 0.15$) APM galaxies
\citep{fosalba:2003}, although the latter has a larger significance,
2.5$\sigma$.  Moreover, the significance of the
detection raises to 3$\sigma$ for the SDSS high-z sample.  The
combined analysis for the 3 samples gives a 3.6$\sigma$ significance
(see \S\ref{sec:chi2}).

Our measurements at large scales are in good agreement with ISW
predictions for a dark-energy dominated universe.  Fig \ref{fig:FitLambda} shows the
probability distribution for $\Omega_\Lambda$ in a flat \lcdm model. We
have fixed $\sigma_8=1$, $h=0.7$ and $\Omega_M+\Omega_\Lambda=1$.  As
we vary $\Omega_\Lambda$ the shape parameter for the linear power
spectrum $P(k)$ consistently changes $\Gamma = h \Omega_M$
\citep{bond:1984}. We only use the data for
$\theta>4^\circ$, where the ISW is the dominant contribution. We fix
the bias $b$ by comparing the matter angular auto-correlation function in
each model to the galaxy auto-correlation in each sample.  We find
$b\simeq 1$ for the APM and {\it SDSS all} samples and $b \simeq
\sqrt{6} $ for the {\it SDSS high-z} sample.  The $\Delta \chi^2$
value in each model refers to the minimum $\chi^2$ fit to a constant
in the range $4^\circ<\theta<10^\circ$.  As can be seen in the figure,
all samples prefer large values of $\Omega_\Lambda$, with the best fit
$\Omega_\Lambda \simeq 0.8$ with a 2$\sigma$ range
$\Omega_\Lambda=0.69-0.87$.

We also see evidence (2.7 $\sigma$ level) for the thermal SZ effect
from the drop of the CMB-galaxy correlation on small-scales in the
low-z samples of SDSS and APM galaxies. These
new measurements can be used to constrain the redshift evolution of the
physical properties of gas inside galaxy clusters.

\acknowledgments

PF wants to thank Francois Bouchet for useful suggestions.
FJC acknowledges useful discussions with Andy Connolly.
We acknowledge support from the Barcelona-Paris bilateral project
(Picasso Programme).  PF acknowledges a post-doctoral CMBNet
fellowship from the EC.  EG and FJC acknowledge the
Spanish MCyT, project AYA2002-00850,
EC-FEDER funding.

\end{document}